\documentclass[3p,times,twocolumn]{elsarticle}
\usepackage{ecrc}
\usepackage{amssymb}
\usepackage{amsmath}
\usepackage{lineno,hyperref}
\modulolinenumbers[5]

\journalname{Nuclear Physics B Proceedings Supplement}
\volume{00}
\firstpage{1}
\runauth{}

\jid{nuphbp}

\jnltitlelogo{Nuclear Physics B Proceedings Supplement}

\begin{document}

\begin{frontmatter}

\title{ Dark matter, dark radiation and Higgs phenomenology \\ in the hidden sector DM models\tnoteref{mytitlenote}}
%\tnotetext[mytitlenote]{Fully documented templates are available in the elsarticle package on %\href{http://www.ctan.org/tex-archive/macros/latex/contrib/elsarticle}{CTAN}.}

\author{Pyungwon Ko\fnref{myfootnote}}
\address{School of Physics, KIAS, Seoul 130-722, Korea}



%\address[mymainaddress]{,Seoul 130-722}

\begin{abstract}
I present a class of hidden sector dark matter (DM) models with local dark gauge symmetries,
where DM is stable due to unbroken local dark gauge symmetry, or due topology, or it is long-lived 
because of some accidental symmetries, and the particle contents and their dynamics are completely 
fixed by local gauge symmetries.  
%It is important to work in renormalizable and unitary models, 
%since we do not know about the mass scales related with the dark sector. 
%where local dark gauge symmetry guarantees stability/longevity of DM particle.
In these models, one have two types of natural force mediators, dark gauge bosons and dark Higgs boson, 
%In general, there are new dark gauge bosons and dark Higgs boson mixing with 
%the SM Higgs boson. These two new particles would
which would affect DM and Higgs phenomenology in important ways.  
I discuss various phenomenological issues including the $\sim$ GeV scale $\gamma$-ray excess from the galactic center (GC), (in)direct detection signatures, dark radiation, Higgs phenomenology and 
Higgs inflation assisted by dark Higgs.
%can be easily accommodated in dark matter models with  local dark gauge symmetry without any conflict 
%with various constraints from direct or indirect detection experiments. 
%Also massless dark gauge boson and/or light dark fermion (sterile neutrinos from  the SM view point) 
%can contribute to dark radiation at a level consistent with Planck data due to Higgs portal interaction. 
%Finally the impact of dark Higgs on the Higgs inflation and the Higgs physics  are discussed. 
\end{abstract}


\end{frontmatter}

%\linenumbers

\section{Introduction}

The standard model (SM) has been tested  from atomic scale up to $\sim O(1)$ TeV scale 
%electroweak scale 
by many experiments, and has been extremely successful. 
However, there are some observational facts which call for new physics beyond the SM 
(BSM): (i) baryon number asymmetry of the universe (BAU), (ii) neutrino masses and mixings, 
(iii) nonbaryonic dark matter (DM) and (iv) inflation in the early universe.

In this talk, I will concentrate on the DM, assuming that BAU and neutrino masses 
and mixings are accommodated by the standard seesaw mechanism by introducing heavy 
right-handed (RH) neutrinos and that the SM  Higgs field drives a successful inflation.   
This talk is based on a series of my works with collaborators ~\cite{Hur:2007uz,Ko:2008ug,Hur:2011sv,Baek:2011aa,Baek:2012uj,Baek:2012se,Baek:2013qwa,Chpoi:2013wga,Baek:2013dwa,Ko:2014nha,Ko:2014bka,Ko:2014gha,Ko:2014eia,Baek:2014jga,Ko:2014loa,Baek:2014kna}. 
For the inflation, I assume that the Higgs inflation is a kind of minimal setup, and 
I show that the dark Higgs from hidden sectors can modify the standard Higgs inflation in a
such a  way that a larger tensor-to-scalar ratio $r \sim O(0.01 - 0.1)$ independent of 
precise values of the top quark and the SM Higgs boson mass \cite{Ko:2014eia}. 
%(i) its tight bond with the top quark and/or the SM Higgs boson mass can be disconnected, 
%and (ii) could lead to a larger tensor-to-scalar ratio $r \sim O(0.01 - 0.1)$ \cite{Ko:2014eia}.  

%%%%%%%%%%%%%%%%%%%%%%%%%%%%%%%%%%%%%%%%%%%%%
\section{Basic assumptions for DM models}

\subsection{Relevant questions for DM} 
So far the existence of DM was confirmed only through the astrophysical and cosmological 
observations where only gravity play an important role.  
%Assuming DM is something similar to the usual particles such electron or proton which are 
described by quantum field theory (QFT), 
We have to seek for the answers to the following questions for better understanding of DM:
%\begin{itemize}
(i) how many species of DM are there in the universe ?
(ii) what are their masses and spins ?
(iii) are they absolutely stable or very long-lived ?
(iv) how do they interact among themselves and with the SM particles ?
(v) where do their masses come from ? 
%\end{itemize}
In order to answer (some of) these questions, we have to observe its signals from colliders 
and/or various (in)direct detection experiments.

%So far, SUSY models have been the (arguably) leading candidate for BSM, because 
%it addresses the fine tuning problem of the Higgs mass, consistent with the idea of grand unification,
%and good CDM candidates (neutralino or graviton LSP).  Since there was no hint for SUSY discovered 
%at the LHC, it would be better to be open-minded about the BSM, especially regarding the new physics
%models regarding the DM. 

The most unique and important property of DM (at least, to my mind) is that DM particle should be 
absolutely stable or long-lived enough, similarly to the case of electron and proton in the SM.  
Let us recall that electron stability is accounted for by electric charge conservation (which is exact), 
and this implies that there should be massless photon, associated with unbroken $U(1)_{\rm em}$ gauge symmetry.   On the other hand, the longevity of proton is ascribed to the baryon number which is an 
accidental global  symmetry of the SM, broken only by dim-6 operators.  
We would like to have DM models where DM is absolutely stable or long-lived enough by similar reasons 
to electron and proton.  And this special property of DM has to be realized in the fundamental Lagrangian 
for DM in a proper way in QFT, similarly to QED and the SM.  Local dark gauge symmetry will play important
roles, by gauranteeing the stability/longevity of DM, as well as determine dynamics in a complete and 
mathematically consistent manner.

\subsection{Hidden sector DM and local dark gauge symmetry}

Any new physics models at the electroweak scale are strongly constrained
by electroweak precision test and CKM phenomenology, if new particles  
feel SM gauge interactions.  The simplest way to evade these two strong constraints is to assume 
a weak scale hidden sector which is made of particles neutral under the SM gauge interaction. 
A hidden sector particle could be a good candidate  for nonbaryonic dark matter of the universe, 
if it is absolutely stable or long lived.  Note that hidden sectors are very generic in many BSMs, including
SUSY models. The hidden sector matters may have their own gauge interactions, which we call 
dark gauge interaction associated with local dark gauge symmetry $G_{\rm hidden}$. 
They can be easily thermalized if there are suitable messengers  between the SM 
and the hidden sectors.  We also assume all the singlet operators such as Higgs portal 
or $U(1)$ gauge kinetic mixing play the role of messengers. 

Another motivation for local dark gauge symmetry $G_{\rm hidden}$ in the hidden 
sector is to stabilize the weak scale DM particle by dark charge conservation laws, 
in the same way electron is absolutely stable because it is the lightest charged 
particle and electric charge is absolutely conserved.  %

Finally note that all the observed particles in Nature feels gauge interactions in addition 
to gravity. 
Therefore it looks very natural to assume that dark matter of the universe (at least some
of the DM species) also feels some (new) gauge force, in addition to gravity.

%%%%%%%%%%%%%%%%%%%%%%%%%%%%%%%%%%%%%%%%%%%%%
\subsection{EFT vs. Renormalizable theories} 

Effective field theory (EFT) approaches are often adopted for DM physics.
For example, let us consider a singlet fermion DM model in EFT: 
\begin{equation}
{\cal L}_{\rm fermion DM}  =   \overline{\psi} \left[  i \not \partial - m_\psi \right] \psi 
- \frac{\lambda_{H\psi}}{\Lambda} H^\dagger H \overline{\psi} \psi 
\end{equation}
with {\it ad hoc} discrete $Z_2$ symmetry under $\psi \rightarrow - \psi$.
However this could be erroneous for a number of reasons.  

Let us consider one of its UV completions \cite{Baek:2011aa}: 
\begin{align}
{\cal L}_{\rm DM} &= {1 \over 2} (\partial_\mu S \partial^\mu S - m_S^2 S^2) 
-\mu_S^3 S - {\mu_S^\prime \over 3} S^3 \nonumber \\\ 
& -  {\lambda_S \over 4} S^4   + \overline{\psi} ( i \not \partial - m_\psi ) \psi 
- \lambda S \overline{\psi} \psi  \nonumber \\\
& -  \mu_{HS} S H^\dagger H -{\lambda_{HS} \over 2} S^2 H^\dag H . 
\label{eq:Lag2}
\end{align}
We have introduced a singlet scalar $S$ in order to make the model (1) renormalizable. 
There will be two scalar bosons $H_1$ and $H_2$ (mixtures of $H$ and $S$) in our model,   
and the additional scalar $S$ makes the DM phenomenology completely different from 
those from Eq. (1).  This is also true for vector DM models~\cite{Baek:2012se,Ko:2014gha}.

For example, the direct detection experiments such as XENON100 and LUX exclude
thermal DM within the EFT model (1), but this is not true within the UV completion 
(2), because of generic cancellation mechanism in the direct detection due to 
a generic destructive interference between $H_1$ and $H_2$ contributions for fermion or vector DM
~\cite{Baek:2011aa,Baek:2012se}.  
Also the direct detection cross section in the UV completion is related with 
that in the EFT  by~\cite{Baek:2014jga}
\begin{equation}
\sigma_{\rm SI}^{\rm ren} = \sigma_{\rm SI}^{\rm EFT} ~
 \left( 1 - \frac{m_{125}^2}{m_1^2} \right)^2~ \cos^4 \alpha \ ,
\end{equation}
which includes the cancellation mechanism and corrects the results reported by ATLAS 
and CMS. 
Here $m_1$ is the mass of the singlet-like scalar boson and $m_{125}$ is the Higgs 
mass found at the LHC. 
Note that the EFT result is recovered when $\alpha \rightarrow 0$ and 
$m_1 \rightarrow \infty$.  

\subsection{Dark Higgs mechanism for the vector DM}

The Higgs portal VDM model is usually described by 
\begin{align}
{\cal L}_{\rm VMD} & = - \frac{1}{4} V_{\mu\nu} V^{\mu\nu} + \frac{1}{2} m_V^2 V_\mu V^\mu  
\nonumber  \\ 
& - \frac{\lambda_{HV}}{2} V_\mu V^\mu |H|^2 - \frac{\lambda_V}{4!} V^4
\end{align}
with an ad hoc $Z_2$ symmetry, $V_\mu \rightarrow -V_\mu$.   Although all the operators are either dim-2 
or dim-4, this Lagrangian breaks gauge invariance, and is neither unitary nor renormalizable. 

One can consider the renormalizable Higgs portal vector DM model by introducing a dark Higgs 
$\Phi$ that generate nonzero mass for VDM by the usual Higgs mechanism: 
\begin{align}
{\cal L}_{VDM} & =  - \frac{1}{4} X_{\mu\nu} X^{\mu\nu} +  
(D_\mu \Phi)^\dagger (D^\mu \Phi)   - \lambda_\Phi \left( | \Phi |^2 - \frac{v_\Phi^2}{2} \right)^2
\nonumber \\
& - \lambda_{\Phi H} \left( | \Phi |^2 - \frac{v_\Phi^2}{2}\right) \left( | H |^2  - \frac{v_H^2}{2}
\right) \ ,
\label{eq:full_theory}
\end{align}
Then the dark Higgs from $\Phi$ mixes with the SM Higgs boson in a similar manner as in SFDM.
And there is a generic cancellation mechanism in the direct detection cross section. Therefore one can
have a wider range of VDM mass compatible with both thermal relic density and direct detection cross 
section  (see Ref.~\cite{Baek:2012se} for more details). In particular the dark Higgs can play an important 
role  in DM phenomenology. 

\begin{figure}
\centering
\includegraphics[scale=0.5]{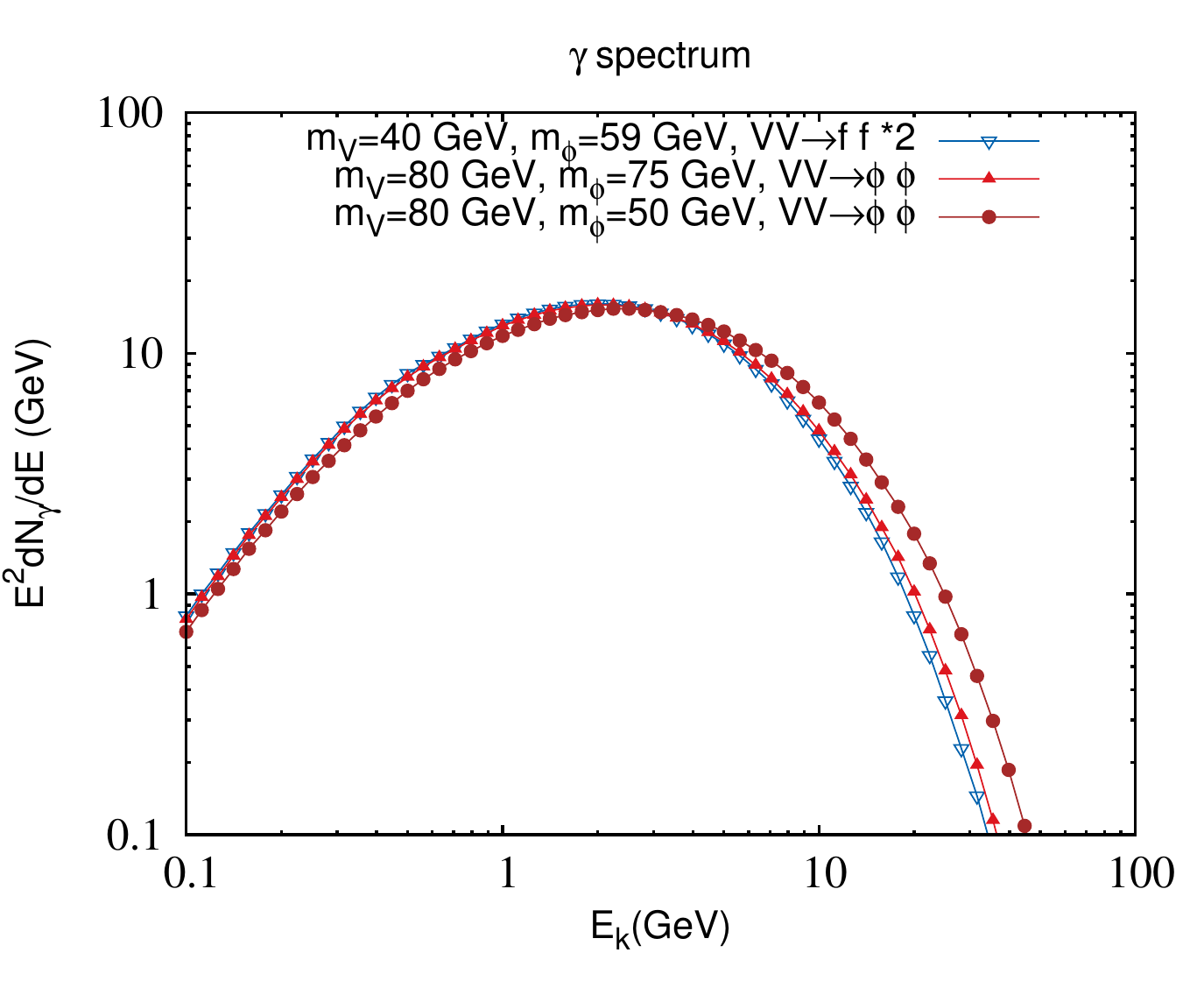}
\caption{Illustration of $\gamma$ spectra from different channels. The first two cases give almost the same spectra while in the third case $\gamma$ is boosted so the spectrum is shifted to higher energy.\label{fig:gammaspectra}}
\end{figure}

%Having the dark Higgs can be very important in DM phenomenology. 
Let me demonstrate it in the context of the GeV scale $\gamma$-ray excess from the galactic center (GC). 
In the Higgs portal VDM with dark Higgs, one can have a new channel for $\gamma$-rays: namely, 
$V V \rightarrow H_2 H_2$ followed by $H_2 \rightarrow b\bar{b}, \tau \bar{\tau}$ through a small mixing
between the SM Higgs and the dark Higgs.  As long as $V$ is slightly heavier than $H_2$ with 
$m_V \sim 80$GeV, one can reproduce the $\gamma$-ray spectrum similar to the one obtained from 
$VV\rightarrow b\bar{b}$ with $m_V \sim 40$GeV (see Fig.~\ref{fig:gammaspectra} and 
Ref.~\cite{Ko:2014gha} for more detail). Note that this mass range for VDM was not allowed within the 
EFT approach based on Eq.~(4),  where there is no room for the dark Higgs at all.  It would have been
simply impossible to accommodate the $\gamma$-ray excess from the galactic center within the 
Higgs portal VDM within EFT. Also this mechanism is generically possible in hidden sector DM models
~\cite{Ko:2014loa}.

%---------------------------------------------------------------------------
%--------------------------------------------------------------------------

%%%%%%%%%%%%%%%%%%%%%%%%%%%%%%%%%%%%%%%%%%%
\section{Stable DM with unbroken dark gauge symmetries: case with local $Z_3$ scalar DM}
%\subsection{Local $Z_3$ scalar DM model}

Let us assume the dark sector has a local $U(1)_{X}$ gauge symmetry spontaneously broken into 
local $Z_{3}$  \'{a} la Krauss and Wilczek.  %~\cite{Krauss:1988zc}. 
This can be achieved with two complex scalar fields $\phi_X$ and $X$   
in the dark sector with the $U(1)_{X}$ charges equal to $1$ and $1/3$, respectively
~\cite{Ko:2014nha,Ko:2014loa}.  Here $\phi_X$ is the dark Higgs that breaks $U(1)_X$ into its
$Z_3$ subgroup by nonzero VEV.  
Then one can write down renormalizable Lagrangian for the SM fields and the dark 
sector fields, $\tilde{X}_\mu, \phi_X$ and $X$: 
\begin{align}
{\cal L} & =  {\cal L}_{{\rm SM}}-\frac{1}{4}\tilde{X}_{\mu\nu}\tilde{X}^{\mu\nu}
-\frac{1}{2}\sin\epsilon\tilde{X}_{\mu\nu}\tilde{B}^{\mu\nu}   \nonumber \\ 
& +D_{\mu}\phi_{X}^{\dagger}D^{\mu}\phi_{X} 
+D_{\mu}X^{\dagger}D^{\mu}X-V (H, X, \phi_X ) \\
V & =  -\mu_{H}^{2} | H |^2 +\lambda_{H} | H^{\dagger}H |^4 -\mu_{\phi}^{2} | \phi_{X} |^2 
+\lambda_{\phi} | \phi_{X} |^4  \nonumber \\ 
&+\mu_{X}^{2} | X |^2 +\lambda_{X} | X |^4 %\nonumber \\
+\lambda_{\phi H} | \phi_{X} |^2 | H |^2   +\lambda_{\phi X} | X |^2 | \phi_{X}|^2  \nonumber \\ 
&+\lambda_{HX} |X|^2 | H |^2 + \left( \lambda_{3}X^{3}\phi_{X}^{\dagger}+H.c. \right)   \label{eq:potential}
\end{align}
where the covariant derivative associated with the gauge field $X^{\mu}$
is defined as $D_{\mu}\equiv\partial_{\mu}-i\tilde{g}_{X}Q_{X}\tilde{X}_{\mu}$. 
%The coupling $\lambda_{3}$ can be chosen as real and positive since it is always possible 
%to  redefine $X$ to absorb the phase.  

We are interested in the phase with the  following vacuum expectation values for the 
scalar fields in the model:
%we are interested in this work is 
\begin{eqnarray}
\langle H\rangle=\frac{1}{\sqrt{2}}\left(\begin{array}{c}
0\\
v_{h}
\end{array}\right),\;\langle\phi_{X}\rangle=\frac{v_{\phi}}{\sqrt{2}},\;\langle X\rangle=0,\label{eq:vacuumstate}
\end{eqnarray}
where only $H$ and $\phi_{X}$ have non-zero vacuum expectation values(vev). This
vacuum will break electroweak symmetry into $U(1)_{\rm em}$, and $U(1)_{X}$ 
symmetry into local $Z_3$, which  stabilizes the scalar field $X$ and make it DM.
%but leave a residual $Z_{3}$ symmetry for $X$. 
The discrete gauge $Z_3$ symmetry stabilizes the scalar DM even if we consider 
higher dimensional nonrenormalizable operators which are invariant under $U(1)_X$.
This is in sharp constrast with the global $Z_3$ model considered in Ref.~\cite{Belanger:2012zr}.
Also the particle contents in local and global $Z_3$ models are different so that the 
resulting DM phenomenology are distinctly different from each other, as summarized in Table 1.

\begin{figure}
\includegraphics[width=0.45\textwidth]{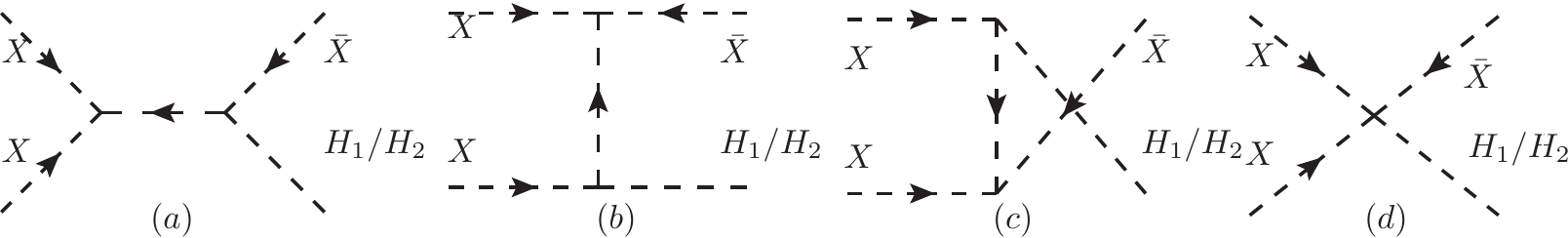}
\includegraphics[width=0.33\textwidth]{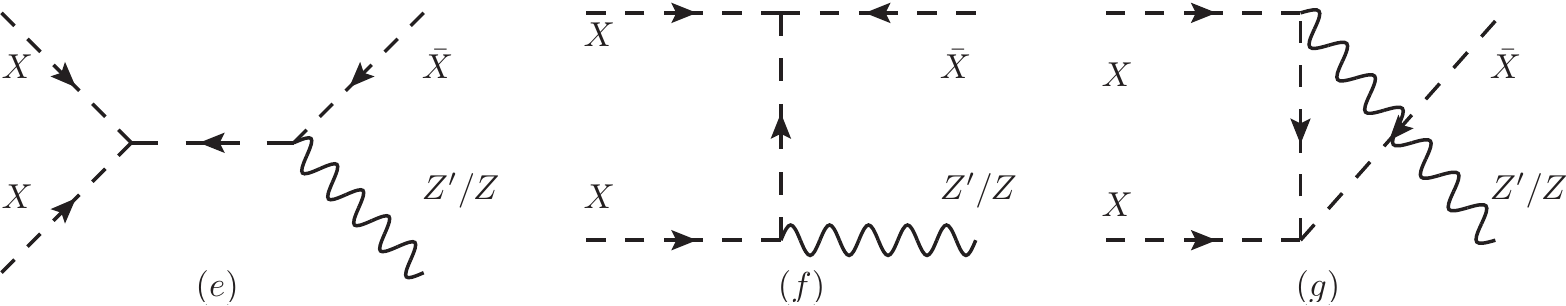} 
\caption{Feynman diagrams for dark matter semi-annihilation. Only (a), (b), and (c) with $H_1$ 
as final state appear in the global $Z_3$ model, while all diagrams could contribute in local $Z_3$ model.
\label{fig:semi-annihilation}}
\end{figure}

\begin{figure}
\includegraphics[width=0.43\textwidth]{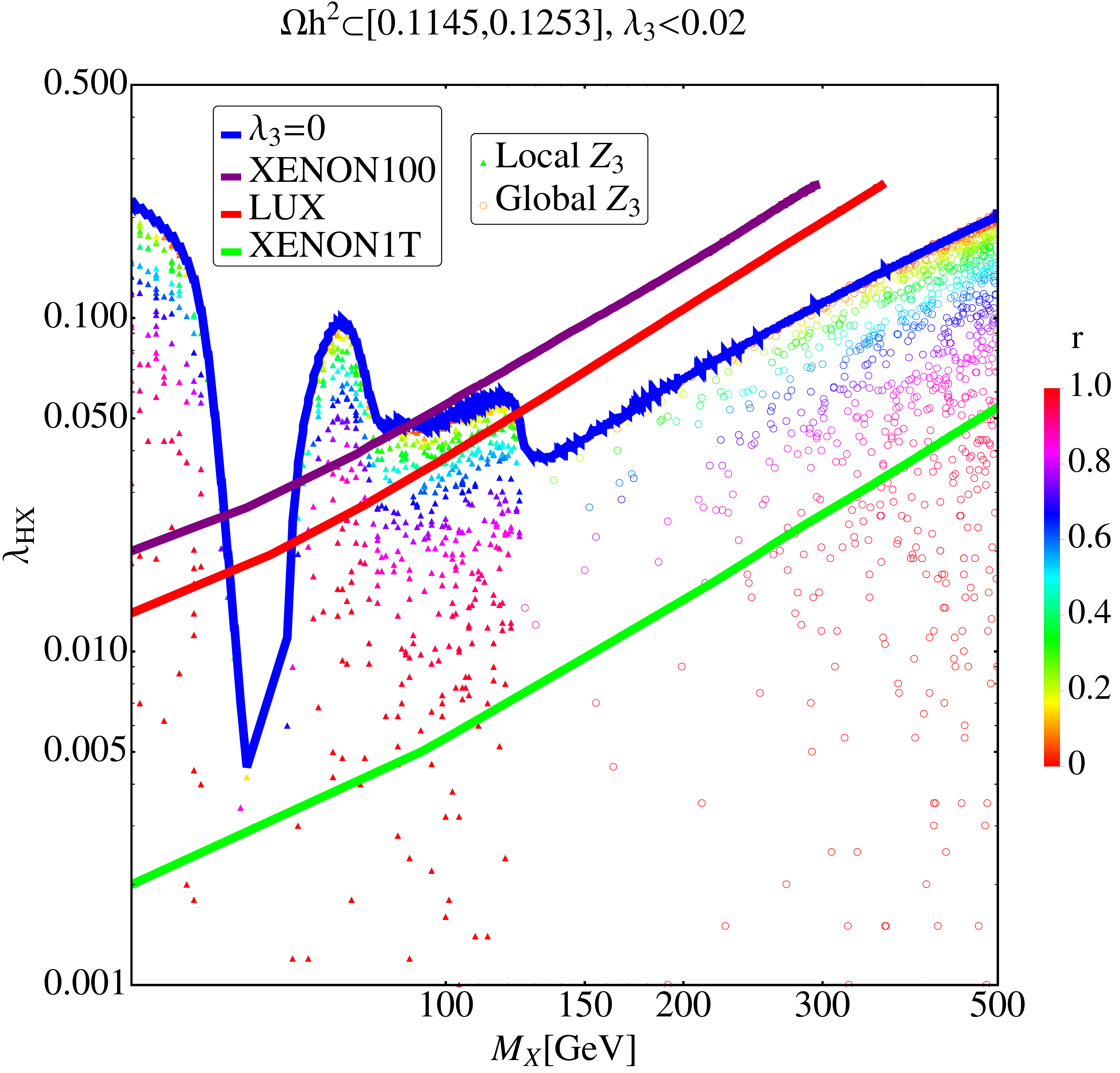}
\caption{Illustration of discrimination between global and local $Z_3$ symmetry. We have chosen 
$M_{H_2}=20$GeV, $M_{Z'}=1$TeV and $\lambda_3<0.02$ as an example. 
%From up to down, three nearly straight lines mark the XENON100~\cite{xenon100}, LUX and expected %XENON1T limits~\cite{xenon1t}, respectively. 
Colors in the scatterred triangles and circles indicate the relative contribution of semi-annihilation, 
$r$ defined in Eq.~(9).  The curved blue band, together with the cirles, gives correct relic density 
of $X$ in the global $Z_3$ model. And the colored triangles appears only in the local $Z_3$ model. 
\label{fig:global_gauge}}
\end{figure}

In Fig.~\ref{fig:semi-annihilation}, I show the Feynman diagrams relevant for thermal relic density 
of local $Z_3$  DM $X$. If we worked in global $Z_3$ DM model instead, we would have diagrams 
only with $H_1$ in (1),(b) and (c).  For local $Z_3$ model, there are two more new fields, dark Higgs 
$H_2$ and dark photon $Z^{'}$, which can make the phenomenology of local $Z_3$ case completely 
difference from that of global $Z_3$ case.  In fact, this can be observed immediately in 
Fig.~\ref{fig:global_gauge}, where the open circles are allowed points in global $Z_3$ model, whereas
the triangles are allowed in local $Z_3$ case. The main difference is that in global $Z_3$ case, the same 
Higgs portal coupling $\lambda_{HX}$ enters both thermal relic density and direct detections.  And 
the stringent constraint from direct detection forbids the region for DM below 120 GeV.  On the other hand
this no longer true in local $Z_3$ case, and there are more options to satisfy all the constraints
~\cite{Ko:2014nha,Ko:2014loa}.

We may define the fraction of the contribution from the semi-annihilation in terms of  
\begin{equation}
\label{eq:r} 
r\equiv\frac{1}{2}\frac{v\sigma^{XX\rightarrow X^{\ast}Y}}{v\sigma^{XX^{\ast}\rightarrow YY}+\frac{1}{2}v\sigma^{XX\rightarrow X^{\ast}Y}}.
\end{equation}

%We can derive the low energy EFT of this model in the limit of very heavy $Z^{'}$ and $H_2$, 
%which would be nothing but the global $Z_3$ model.   However, if we started from global $Z_3$ model 
%from the beginning with higher dimensional operators, the stability of DM $X$ would not be guaranteed 
%in general.
Also one can drive the low energy EFT and discuss its limitation, the details
of which can be found in Ref.~\cite{Ko:2014nha}.  The main message is that the EFT cannot enjoy 
the advantages of having the full particles spectra in the gauge theories, namely not-so-heavy dark Higgs
and dark gauge bosons, which could be otherwise helpful for explaining the galactic center 
$\gamma$-ray excess or the strong self-interacting DM.  
And it is important to know what symmetry stabilizes the DM particles.

\begin{table}[htdp]
\begin{center}
\begin{tabular}{|c||c|c|}
\hline
           & Global $Z_3$ & Local $Z_3$   \\ \hline  \hline% \\
Extra fields &  $X$ & $X, Z^{'}, \phi$    \\   \hline
Mediators & $H$ & $H$, $Z^{'}$, $\phi$    \\   \hline
Constraints & Direct detection   & Can be relaxed \\
                   & Vacuum stability &  Can be relaxed \\    \hline
DM mass & $m_X \gtrsim 120$GeV  & $m_X < m_H$ allowed     
\\ 
\hline  
\end{tabular}
\end{center}
\caption{\label{tab:z3} 
Comparison between the global and the local $Z_3$ scalar dark matter models.
Here $X$ is a complex scalar DM,  $H$ is the observed SM-HIggs like boson, and $\phi$ is the dark 
Higgs from $U(1)_X$  breaking into $Z_3$ subgroup. 
}
\end{table}

%Now let us make a brief comparison of the local $Z_3$ scalar DM model 
%with the effective field theory (EFT) approach.   For a complex scalar DM $X$ with local $Z_3$, 
%one can easily construct the following operators imposing $Z_3$ symmetry,  to list only a few:
%\begin{align}
%U(1)_X ~{\rm sym} : & X^\dagger X H^\dagger H ,  \  
%%\frac{1}{\Lambda^2} \left( X^\dagger D_\mu X \right) 
%%\left( H^\dagger D^\mu H \right) , \    \nonumber \\
%%& 
%\frac{1}{\Lambda^2} \left( X^\dagger D_\mu X \right) 
%\left( \overline{f} \gamma^\mu f \right)  
%\\
%%
%\frac{1}{\Lambda} X^3 H^\dagger H , \ \  \frac{1}{\Lambda^3} X^3  \overline{f_L} H f_R  
%\end{align}
%where $f$ is a SM fermion field and $\Lambda$ is a combination of new physics scale
%and couplings of the DM particle to new physics particle, and can differ from one operator
%to another. Note that the operators in Eq.~(12) relevant for semi-annihilation are usually missed 
%in the operator analysis, since it was unique only in $Z_3$ (local or global) dark matter models.  
%Our arguments show that it is important to implement properly the dark symmetry which stabilize 
%the DM particles.  Depending on dark gauge symmetries, we will have different (effective) operators.  
%%The usual story within the EFT is that the direct detection cross section due to the renormalizable 
%%operator $X^\dagger X H^\dagger H$ is strongly constrained so that  the scalar DM can not be 
%thermalized if it is light.

Local $Z_2$ dark matter is also another interesting case, and easily compared with the usual 
global $Z_2$ scalar DM model.  It would be have similar aspects as in local $Z_3$ model, and 
the detailed discussions can be found in Ref.~\cite{Baek:2014kna}.  Also one can consider unbroken
$U(1)_X$ dark gauge symmetry with scalar DM and the RH neutrinos decay both to the SM and the dark 
sector particles.  See Ref.~\cite{Baek:2013qwa} for more details. 

\section{Stable DM due to topology: Hidden sector monopole and vector DM, dark radiation}

In field theory there could be a topologically stable classical configuration.  The most renowned 
example is the 't Hooft-Polyakov monopole. This object in fact puts a serious problem in cosmology,
and was one of the motivations for inflationary paradigm. 
In Ref.~\cite{Baek:2013dwa}, we revived this noble idea by putting the monopole in the hidden 
sector and introducing the Higgs portal interaction to connect the hidden and the visible sectors.

Let us consider $SO(3)_X$-triplet real scalar field $\vec{\Phi}$ with the following Lagrangian implemented 
to the SM: % Lagrangian:
\begin{align} \label{Lag}
{\cal L}_{\rm new} & = - \frac{1}{4} V_{\mu\nu}^a V^{a \mu\nu} + \frac{1}{2} 
D_\mu \vec{\Phi} \cdot D^\mu \vec{\Phi}   \nonumber  \\
& -  \frac{\lambda_\Phi}{4} 
\left( \vec{\Phi} \cdot \vec{\Phi} - v_\phi^2 \right)^2  \nonumber  \\
& - \frac{\lambda_{\Phi H}}{2}   \left(\vec{\Phi} \cdot \vec{\Phi} - v_\phi^2\right)
\left( H^\dagger H - \frac{v_H^2}{2} \right) .
\end{align}
%where ${\cal L}_{\rm SM}$ is the standard model Lagrangian $D_\mu \Phi^a = \partial_\mu \Phi^a - g_X  %\epsilon^{abc} V_\mu^b \Phi^c$ and $V_{\mu\nu}^a = \partial_\mu V_\nu^a - \partial_\nu V_\mu^a - g_X  \epsilon^{abc} V_\mu^b V_\nu^c$ with $g_X$ and $\epsilon^{abc}(a,b,c=1,2,3)$ being the gauge 
%coupling and the structure constant of the hidden sector $SO(3)_X \approx SU(2)_X$ gauge group. 
The Higgs portal interaction is described by the $\lambda_{\Phi H}$ term, which is a new addition to the 
renowned 't Hooft-Polyakov monopole model.  %~\cite{'tHooft:1974qc,Polyakov:1974ek}.
%When we ignore the Higgs portal interaction, the hidden sector Lagrangian describes 
%the famous 't Hooft-Polyakov monopole~\cite{'tHooft:1974qc,Polyakov:1974ek}.

After the spontaneous symmetry breaking of $SO(3)_X$ into $SO(2)_X (\approx U(1)_X)$ by nonzero 
vacuum expectation value (VEV) of $\vec{\Phi}$ with  $\langle \vec{\Phi} (x) \rangle = ( 0 , 0 , v_\Phi )$, 
hidden sector particles are composed of massive dark vector bosons $V_\mu^\pm$ 
\footnote{Here $\pm 1$ in $V_\mu^\pm$ indicate the dark charge 
under $U(1)_X$, and not ordinary electric charges.} with masses $m_V=g_X v_\Phi$ (which are stable 
due to the unbroken subgroup $SO(2)_X \approx U(1)_X$),  massless dark photon 
$\gamma_{h, \mu} \equiv V_\mu^3$,  topologically stable heavy (anti-)monopole  with mass 
$m_M \sim m_V/\alpha_X$, and massive real scalar  $\phi$ (dark Higgs boson) mixed with the SM 
Higgs boson through the Higgs portal term. 
%Also, after the spontaneous breaking of electroweak symmetry, Higgs portal interaction mixes 
%$\phi$ and SM Higgs boson $h$, $h$ and $\phi$, similarly to the renormalizable models for singlet fermion 
%DM \cite{fermion_dm,vacuum_stability}   or VDM~\cite{hambye,vector_dm}.  

Note that there is no kinetic mixing between $\gamma_h$ and the SM {$U(1)_Y$-gauge boson unlike  
the $U(1)_X$-only case,   due to the non Abelian nature of the hidden gauge symmetry.  Also the VDM is 
stable even in the presence of nonrenormalizable operators due to the unbroken subgroup $U(1)_X$. 
This would not have been the case, if the $SU(2)_X$ were completely broken by a complex 
$SU(2)_X$ doublet,  where the stability of massive VDM is not protected by $SU(2)_X$ gauge symmetry 
and nonrenormalizable  interactions would make the VDM  decay in general~\cite{Hambye:2008bq}.  
Of course, it would be fine 
as long as the lifetime of the decaying VDM is long enough so that it can  still be a good CDM candidate.
In the VDM model with a hidden sector monopole, the unbroken $U(1)_X$ subgroup not only protects 
the stability of VDM $V_\mu^\pm$, but also contributes to the dark radiation at the level of $\sim 0.1$.  
We refer the readers to the original paper on more details of phenomenology of this model
~\cite{Baek:2013dwa} (see also Ref.~\cite{khoze}).

%%%%%%%%%%%%%%%%%%%%%%%%%%%%%%%%%%%%%%%%%%%%%%
\section{EWSB and CDM from Strongly Interacting Hidden Sector:   
long-lived DM due to accidental symmetries} 
%%%%%%%%%%%%%%%%%%%%%%%%%%%%%%%%%%%%%%%%%%%%%%

Another nicety of models with hidden sector is that one can construct a model 
where all the mass scales of the SM particles and DM are generated by 
dimensional transmutation in the hidden sector~\cite{Hur:2007uz,Ko:2008ug,Hur:2011sv}. 
Basically the light hadron masses such as proton or $\rho$ meson come from confinement, which is 
derived from massless QCD through dimensional transmutation.  One can ask if all the masses of 
observed particles can be generated by quantum mechanics, in a similar manner with the proton mass
in the massless QCD.  The most common way to address this question is to employ the Coleman-Weinberg
mechanism for radiative symmetry breaking.  Here I present a new model based on nonperturbative 
dynamics  like technicolor or chiral symmetry breaking in ordinary  QCD.

Let us consider a scale-invariant extension of the SM with a strongly interacting hidden sector: 
\begin{align}
  {\cal L}  & =  {\cal L}_{\rm SM, kin} + {\cal L}_{\rm SM, Yukawa}
- {\lambda_{H} \over 4}~( H^{\dagger} H )^2 \nonumber \\
& - {\lambda_{SH} \over 2}~S^2 ~ H^{\dagger} H - {\lambda_S \over 4}~S^4 
- \frac{1}{4} {\cal G}_{\mu\nu}^a {\cal G}^{a \mu\nu}  \nonumber \\
& + \sum_{k=1,...,f} \overline{\cal Q}_k \left[   i D \cdot \gamma  - \lambda_k S 
\right] {\cal Q}_k .
\end{align}
Here ${\cal Q}_k$ and ${\cal G}_{\mu\nu}^a$ are the hidden sector quarks and gluons, and 
and the index $k$ is the flavor index in the hidden sector QCD. 
In this model, we have assumed that the hidden sector strong interaction is vectorlike and confining 
like the ordinary QCD.  Then we can use the known aspects of QCD dynamics to the hidden sector QCD.

Note that the real singlet scalar $S$  plays the role of messenger connecting 
the SM Higgs sector and the  hidden sector quarks.  

In this model, dimensional transmutation in the hidden sector will generate the
hidden QCD scale and chiral symmetry breaking with developing
nonzero $\langle \bar{\cal Q}_k {\cal Q}_k \rangle$. 
 Once  a nonzero $\langle \bar{\cal Q}_k {\cal Q}_k \rangle$ is developed, 
the  $\lambda_k S $ term generate the linear potential for the real  singlet $S$, 
leading to nonzero $\langle S \rangle$. This in turn generates the hidden sector current 
quark masses through $\lambda_k$ terms as well as the EWSB through $\lambda_{SH}$ term.
Then the Nambu-Goldstone boson $\pi_h$ will get nonzero masses, and becomes 
a good CDM  candidate. Also hidden sector baryons ${\cal B}_h$ will be formed, 
the lightest of which would be long lived due to the accidental h-baryon conservation.
See Ref.~\cite{Hur:2011sv} for more details.
%The messenger $S$ can thermalize efficiently $\pi_h$'s and lead to direct detection 
%cross section at the observable level.  
%pair  annihilation into the SM  particles occurs more efficiently, and it is easy to 
%accommodate the WMAP  data on $\Omega_{\rm CDM} h^2$. Direct detection rates of 
%$\pi_h$ are in the interesting ranges.  

%%%%%%%%%%%%%%%%%%%%%%%%%%%%%%%%%%%%%%%%%%%%%%%%
\section{Light mediators and Self-interacting DM}
%%%%%%%%%%%%%%%%%%%%%%%%%%%%%%%%%%%%%%%%%%%%%%%%

Another nice feature of the dark matter models with local dark gauge symmetry is that the model 
includes new degrees of freedom,  dark gauge bosons  and dark Higgs boson(s), that can play 
the role of force mediators from the beginning because of the rigid structure of the underlying gauge 
theories.  
%Namely one always have dark gauge bosons  and dark Higgs boson(s), whose masses are not known.  
In fact one can utilize the light mediators in order to explain the GeV scale $\gamma$-ray excess 
or the self-interacting DM which would solve three puzzles in the CDM paradigm: 
(i) core-cusp problem, (ii) missing satellite problem and (iii) too-big-to-fail problem.
These would have been simply impossible if we adopted the EFT approach for DM physics.

In the EFT approach for the DM, these new degrees of freedom are very heavy compared with the 
DM mass as well as the energy scale we are probing the dark sector (e.g., the collider energy scale).  
However, we don't know anything about the mass scales of these mediators, and it would be too strong 
an assumption. Without these light mediators, we could not explain the GeV scale $\gamma$-ray excess
as described in this talk, or have strong self-interacting DM.  This illustrates one of the limitations of DM 
EFT appraoches. 

%%%%%%%%%%%%%%%%%%%%%%%%%%%%%%%%%%%%%%%%%%%%%%%%
\section{Higgs inflation assisted by the Higgs portal }
%%%%%%%%%%%%%%%%%%%%%%%%%%%%%%%%%%%%%%%%%%%%%%%%

The final issue related with DM models with local dark gauge symmetris is the Higgs inflation 
in the presence of the Higgs portal interaction to the dark sector:   
%From the previous discussions, it is clear that  there will be a dark Higgs from dark sectors, 
%and it will have the Higgs portal interactions in most cases. 
%This dark Higgs makes another strong motivation for singlet scalar extensions of the SM. 
%Let us remind ourselves that singlet scalar extensions have been discussed mostly  in the context 
%of the tree level $\rho$-parameter, and the strong 1st order phases transition for electroweak baryogenesis. 
%On the other hand, I showed that the dark Higgs boson which is (almost) ubiquitous in  DM models
%with local dark symmetries and/or renormalizable Higgs portal DM models plays just the right role
%as a singlet scalar from the SM view point. If we wish to understand the absolute stability or longevity 
%of the weak scale DM in terms of local dark gauge symmetry, we will always have a guest ``dark Higgs = 
%singlet scalar''.
\begin{equation} \label{L-jordan}
\frac{\mathcal{L}}{\sqrt{-g}} = - \frac{1}{2 \kappa} \left( 1 + \xi \frac{h^2}{M_{\rm Pl}^2} 
\right) R + \mathcal{L}_h  + \lambda_{\phi H} \phi^2 h^2  
\end{equation}
in the unitary gauge, 
where $\kappa = 8 \pi G = 1/M_{\rm Pl}^2$ with $M_{\rm Pl}$ being the reduced Planck mass, 
and $\mathcal{L}_h$ is the Lagrangian of the SM Higgs field only.    Here $\phi$ denotes a generic 
dark Higgs field which mixes with the SM Higgs field after dark and EW gauge symmetry breaking. 
In the presence of the Higgs portal interaction, we have recalculated the slow-roll parameters.
% and found that it is possible to have the correct value
%of $n_s$ with small spectral running, and $\sim O(0.1)$ ~\cite{Ko:2014eia}.  
Relegating the details to Ref.~\cite{Ko:2014eia},  I simply show the results: 
%here I show how the Jordan-frame Higgs potential ($V_h$) depends on the mixing angle 
%$\alpha$ for fixed values of $m_\phi$ and $\lambda_{\Phi H}$ (Fig.~\ref{fig:Vhiggs}).  
%---------------------
%It is clear that for a given $m_\phi$ the inflection point is determined by 
%delicate interplay between $\alpha$ and $\lambda_{\Phi H}$. 
%Note that  one can achieve  nearly the same behavior of the Higgs potential 
%by adjusting $m_\phi$ instead of $\alpha$.
%I show the inflaton potential originating from $V_h$ with an inflection point in Fig.~\ref{fig:UhiggsRG}.
%---------------------
%\begin{figure}[h]
%\centering
%\includegraphics[width=0.48\textwidth]{Uhiggs-RGrun-inflextion.eps}
%\caption{\label{fig:UhiggsRG}
%$U(h)$ of Eq.~(\ref{Uhiggs-RG}) in in SFDM for $m_h=125$GeV, $M_t=173.2$GeV, $m_\phi = 500$GeV, 
%$\lambda_{SH} = 0.055$, $\lambda_S = 0.2$, and $\lambda_\psi = 0.4$ and $\alpha = 0.074222$.
%}
%\end{figure}
%---------------------
%At the benchmark point used in Fig.~\ref{fig:UhiggsRG}, 
%$V_I \sim 9 \times 10^{64} {\rm GeV}^4$ 
%during inflation.  Since radiation dominates the universe right after inflation in this scenario with 
%$\xi \sim 10$ \cite{Bezrukov:2014bra} 
%\footnote{For $\xi \sim 10$, the issue of perturbative unitarity 
%becomes less severe.}, one finds
%$T_{\rm R} \sim V_I^{1/4} \sim 1.7 \times 10^{16}$GeV.
%Based on this, numerical analysis was performed 
at a bench mark point for Fig.~2 of Ref.~\cite{Ko:2014eia}, 
%with $m_h=125$GeV, $M_t=173.2$GeV, $m_\phi = 500$GeV, 
%$\lambda_{SH} = 0.055$, $\lambda_S = 0.2$, and $\lambda_\psi = 0.4$ and $\alpha = 0.074222$
%for a pivot scale $k_*=0.05 {\rm Mpc}^{-1}$,   
we get the following results: 
\begin{equation}
n_s = 0.9647 \ , \ r = 0.0840 \ , \  
\end{equation}
for $N_e = 56$, $h_* / M_{\rm Pl} = 0.72$, $\alpha =  0.07422199$ and $\xi = 12.8294$ 
for a pivot scale $k_*=0.05 {\rm Mpc}^{-1}$.
There is a parameter space where the spectral running of $n_s$ is small enough at the level of
$|  n_s^{'} | \lesssim 0.01$.  It is amusing to notice that the $r$ could be as large as 
$\sim O(0.1)$ in the presence of the Higgs portal interactions to a dark sector, independent of 
the top quark and the Higgs boson mass in the standard Higgs inflation scenario.

%%%%%%%%%%%%%%%%%%%%%%%%%%%%%%%%%%%%%%%%%%%%%%%%
%\section{Summary of  hidden sector DM models}
\section{Higgs phenomenology, EW vacuum stability,  and dark radiation}
%%%%%%%%%%%%%%%%%%%%%%%%%%%%%%%%%%%%%%%%%%%%%%%%
%Here we summarize generic aspects of hidden sector DM models with local dark 
%gauge symmetries, independent of dark gauge symmetries. 
%\begin{itemize}
%\item 
Now let us discuss Higgs phenomenology within this class of DM models. 
Due to the mixing effect between the dark Higgs and the SM Higgs bosons, 
the signal strengths of the observed Higgs boson will be universally reduced from ''1''  
independent of production and decay channels~\cite{Baek:2011aa,Baek:2012se}. 
Also the 125 GeV Higgs boson could decay into a pair of dark Higgs and/or a pair of dark gauge boson, 
which is still  allowed by the current LHC data~\cite{Chpoi:2013wga}. These predictions 
will be further constrained by the next round experiments. 

Also the dark Higgs can make the EW vacuum stable upto the Planck scale without 
any other new physics~\cite{Baek:2012uj,Baek:2012se}, and this was very important in the 
Higgs-portal assisted Higgs inflation discussed in the previous  section.
%\item Direct detection constraints on the DM model are relaxed due to the cancellation
%mechanism for the fermion or vector DM~\cite{Baek:2011aa,Baek:2012uj}.
%\item Thermal relic density of DM can be modified due to new channels if dark Higgs and/or
%dark gauge bosons lighter than the DM: $DM + DM \rightarrow \phi \phi , \gamma^{'} \gamma^{'}$. 
%Thus a wider range of DM mass would be compatible with thermal DM.  For example,  
%VDM model with $m_{\rm VDM} \sim 80$ GeV and $m_{\phi} \sim 75$ GeV can accommodate the 
%$\gamma$-ray excess from the galactic center, which would be impossible within the EFT 
%VDM model~\cite{Ko:2014gha}.
%\item The standard Higgs inflation is modified through the dark Higgs mass parameter and 
%the mixing angle $\alpha$, and can accommodate a larger $r \sim O(0.1)$ 
%(tensor-to-scalar ratio) compared with the original Higgs inflation~\cite{Ko:2014eia}. 

In most cases, there is generically a singlet scalar which is nothing but a dark Higgs, which would give 
a new motivation to consider singlet extensions of the SM. Traditionally a singlet scalar was 
motivated mainly  by why-not or $\Delta \rho$ constraint, or the strong first order EW phase transition for 
electroweak baryogenesis.
Being a singlet scalar, the dark Higgs will satisfy all these motivations, as well as stability of DM 
by local dark gauge symmetry.   
It would be important to seek for this singlet-like scalar at the LHC or the ILC, but the colliders 
cannot cover the entire mixing angle down to $\alpha \sim 10^{-6}$ relevant to DM phenomenology. 

Massless dark gauge boson or light dark fermions in hidden sectors could 
contribute to dark radiation of the universe  
In a class of models we constructed, the amount of extra dark radiation is rather small
by an amount consistent with the Planck data due to Higgs portal  interactions
~\cite{Baek:2013qwa,Baek:2013dwa,Ko:2014bka}.
%\end{itemize}

%%%%%%%%%%%%%%%%%%%%%%%%%%%%%%%%%%%%%%%%%%%%%%
%\section{Conclusion}
%In this talk, I described a class of weak DM model where local dark gauge symmetry and 
%renormalizability play crucial roles both in theoretical consistency and phenomenology.

%%%%%%%%%%%%%%%%%%%%%%%%%%%%%%%%%%%%%%%%%%%%%%

\section{Acknowledgements}

The author is grateful to G. Ricciardi and M. Neubert for organizing this workshop and the 
support. He also thanks S. Baek, T. Hur, D.W.Jung, W.I. Park, E. Senaha, Y. Tang 
for enjoyable collaborations on the subjects presented in this talk.
This work is supported in part by the NRF Research Grant 2012R1A2A1A01006053.


\begin{thebibliography}{999}

%\cite{Hur:2007uz}
\bibitem{Hur:2007uz}
  T.~Hur, D.~W.~Jung, P.~Ko and J.~Y.~Lee,
  %``Electroweak symmetry breaking and cold dark matter from strongly interacting hidden sector,''
  Phys.\ Lett.\ B {\bf 696} (2011) 262. 
%  [arXiv:0709.1218 [hep-ph]]; 
  %%CITATION = ARXIV:0709.1218;%%
  
%\cite{Ko:2008ug}
\bibitem{Ko:2008ug}
  P.~Ko,
  %``Electroweak symmetry breaking and cold dark matter from hidden sector technicolor,''
  Int.\ J.\ Mod.\ Phys.\ A {\bf 23} (2008) 3348.  
%  [arXiv:0801.4284 [hep-ph]]; 
  %%CITATION = ARXIV:0801.4284;%%
  
%\cite{Hur:2011sv}
\bibitem{Hur:2011sv}
  T.~Hur and P.~Ko,
  %``Scale invariant extension of the standard model with strongly interacting hidden sector,''
  Phys.\ Rev.\ Lett.\  {\bf 106} (2011) 141802.
%  [arXiv:1103.2571 [hep-ph]].
  %%CITATION = ARXIV:1103.2571;%%

%\cite{Baek:2011aa}
\bibitem{Baek:2011aa}
  S.~Baek, P.~Ko and W.~I.~Park,
  %``Search for the Higgs portal to a singlet fermionic dark matter at the LHC,''
  JHEP {\bf 1202} (2012) 047.
%  [arXiv:1112.1847 [hep-ph]].
  %%CITATION = ARXIV:1112.1847;%%

%\cite{Baek:2012uj}
\bibitem{Baek:2012uj}
  S.~Baek, P.~Ko, W.~I.~Park and E.~Senaha,
  %``Vacuum structure and stability of a singlet fermion dark matter model with a singlet scalar messenger,''
  JHEP {\bf 1211} (2012) 116.
%  [arXiv:1209.4163 [hep-ph]].
  %%CITATION = ARXIV:1209.4163;%%

%\cite{Baek:2012se}
\bibitem{Baek:2012se}
  S.~Baek, P.~Ko, W.~I.~Park and E.~Senaha,
  %``Higgs Portal Vector Dark Matter : Revisited,''
  JHEP {\bf 1305} (2013) 036.
%  [arXiv:1212.2131 [hep-ph]].
  %%CITATION = ARXIV:1212.2131;%%

%\cite{Baek:2013qwa}
\bibitem{Baek:2013qwa}
  S.~Baek, P.~Ko and W.~I.~Park,
  %``Singlet Portal Extensions of the Standard Seesaw Models to a Dark Sector with Local Dark Symmetry,''
  JHEP {\bf 1307} (2013) 013.
%  [arXiv:1303.4280 [hep-ph]].
  %%CITATION = ARXIV:1303.4280;%%

%\cite{Chpoi:2013wga}
\bibitem{Chpoi:2013wga}
  S.~Choi, S.~Jung and P.~Ko,
  %``Implications of LHC data on 125 GeV Higgs-like boson for the Standard Model and its various extensions,''
  JHEP {\bf 1310} (2013) 225.
%  [arXiv:1307.3948].
  %%CITATION = ARXIV:1307.3948;%%

%\cite{Baek:2013dwa}
\bibitem{Baek:2013dwa}
  S.~Baek, P.~Ko and W.~I.~Park,
  %``Hidden sector monopole, vector dark matter and dark radiation with Higgs portal,''
  JCAP {\bf 1410} (2014) 067.
%  arXiv:1311.1035 [hep-ph].
  %%CITATION = ARXIV:1311.1035;%%

%\cite{Ko:2014nha}
\bibitem{Ko:2014nha}
  P.~Ko and Y.~Tang,
  %``Self-interacting scalar dark matter with local $Z_3$ symmetry,''
  JCAP {\bf 1405} (2014) 047.
%  [arXiv:1402.6449 [hep-ph], arXiv:1402.6449].
  %%CITATION = ARXIV:1402.6449;%%

%\cite{Ko:2014bka}
\bibitem{Ko:2014bka}
  P.~Ko and Y.~Tang,
  %``$\nu \Lambda$MDM: A Model for Sterile Neutrino and Dark Matter Reconciles Cosmological and Neutrino Oscillation Data after BICEP2,''
   Phys.Lett. {\bf B739} (2014) 62.
%  arXiv:1404.0236 [hep-ph].
  %%CITATION = ARXIV:1404.0236;%%

%\cite{Ko:2014gha}
\bibitem{Ko:2014gha}
  P.~Ko, W.~I.~Park and Y.~Tang,
  %``Higgs portal vector dark matter for $\mathinner{\mathrm{GeV}}$ scale $\gamma$-ray excess from galactic center,''
  JCAP {\bf 1409} (2014) 013.
%  [arXiv:1404.5257 [hep-ph]].
  %%CITATION = ARXIV:1404.5257;%%

%\cite{Ko:2014eia}
\bibitem{Ko:2014eia}
  P.~Ko and W.~I.~Park,
  %``Higgs-portal assisted Higgs inflation in light of BICEP2,''
  arXiv:1405.1635 [hep-ph].
  %%CITATION = ARXIV:1405.1635;%%

%\cite{Baek:2014jga}
\bibitem{Baek:2014jga}
  S.~Baek, P.~Ko and W.~I.~Park,
  %``Invisible Higgs Decay Width vs. Dark Matter Direct Detection Cross Section in Higgs Portal Dark Matter Models,''
  Phys.\ Rev.\ D {\bf 90} (2014) 055014.
%  [arXiv:1405.3530 [hep-ph]].
  %%CITATION = ARXIV:1405.3530;%%

%\cite{Ko:2014loa}
\bibitem{Ko:2014loa} 
  P.~Ko and Y.~Tang,
  %``Galactic center $\gamma$-ray excess in hidden sector DM models with dark gauge symmetries: local $Z_{3}$ symmetry as an example,''
  JCAP {\bf 1501} (2015) 023.
%  arXiv:1407.5492 [hep-ph].
  %%CITATION = ARXIV:1407.5492;%%
  
%\cite{Baek:2014kna}
\bibitem{Baek:2014kna}
  S.~Baek, P.~Ko and W.~I.~Park,
  %``Global vs. Local $Z_2$ Symmetries for Real Scalar Dark Matter,''
  arXiv:1407.6588 [hep-ph].
  %%CITATION = ARXIV:1407.6588;%%
  %2 citations counted in INSPIRE as of 12 Oct 2014

%\cite{Belanger:2012zr}
\bibitem{Belanger:2012zr}
  G.~Belanger, K.~Kannike, A.~Pukhov and M.~Raidal,
  %``$Z_3$ Scalar Singlet Dark Matter,''
  JCAP {\bf 1301} (2013) 022.
%  [arXiv:1211.1014 [hep-ph]].
  %%CITATION = ARXIV:1211.1014;%%

%\cite{Hambye:2008bq}
\bibitem{Hambye:2008bq}
  T.~Hambye,
  %``Hidden vector dark matter,''
  JHEP {\bf 0901} (2009) 028.
%  [arXiv:0811.0172 [hep-ph]].
  %%CITATION = ARXIV:0811.0172;%%

%\cite{Khoze:2014woa}
\bibitem{khoze}
  V.~V.~Khoze and G.~Ro,
  %``Dark matter monopoles, vectors and photons,''
  JHEP {\bf 1410} (2014) 61.
%  [arXiv:1406.2291 [hep-ph]].
  %%CITATION = ARXIV:1406.2291;%%
\end{thebibliography}
\end{document}